\documentclass[aps,pra,twocolumn,superscriptaddress]{revtex4}
\usepackage{amsmath,amsfonts,amssymb}
\usepackage{color,graphicx,hyperref,ulem}

\begin{document}
\title{Divide-and-Conquer: An integrated photon-counting scheme}

\author{Ren\'{e} Heilmann}\affiliation{Institute of Applied Physics, Abbe Center of Photonics, Friedrich-Schiller-Universit\"at Jena, Max-Wien-Platz 1, 07743 Jena, Germany}
\author{Jan Sperling}\affiliation{Arbeitsgruppe Theoretische Quantenoptik, Institute f\"ur Physik, Universit\"at Rostock, 18051 Rostock, Germany}
\author{Armando Perez-Leija}\affiliation{Institute of Applied Physics, Abbe Center of Photonics, Friedrich-Schiller-Universit\"at Jena, Max-Wien-Platz 1, 07743 Jena, Germany}
\author{Markus Gr\"{a}fe}\affiliation{Institute of Applied Physics, Abbe Center of Photonics, Friedrich-Schiller-Universit\"at Jena, Max-Wien-Platz 1, 07743 Jena, Germany}
\author{Matthias Heinrich}\affiliation{Institute of Applied Physics, Abbe Center of Photonics, Friedrich-Schiller-Universit\"at Jena, Max-Wien-Platz 1, 07743 Jena, Germany}
\author{Stefan Nolte}\affiliation{Institute of Applied Physics, Abbe Center of Photonics, Friedrich-Schiller-Universit\"at Jena, Max-Wien-Platz 1, 07743 Jena, Germany}
\author{Werner Vogel}\affiliation{Arbeitsgruppe Theoretische Quantenoptik, Institute f\"ur Physik, Universit\"at Rostock, 18051 Rostock, Germany}
\author{Alexander Szameit}\email{alexander.szameit@uni-jena.de}\affiliation{Institute of Applied Physics, Abbe Center of Photonics, Friedrich-Schiller-Universit\"at Jena, Max-Wien-Platz 1, 07743 Jena, Germany}

\begin{abstract}
	The key requirement for harnessing the quantum properties of light is the capability to detect and count individual photons.
	Of particular interest are photon-number-resolving detectors, which allow one to determine whether a state of light is classical or genuinely quantum.
	Existing schemes for addressing this challenge rely on a proportional conversion of photons to electrons.
	As such, they are capable of correctly characterizing small photon fluxes, yet are limited by uncertainties in the conversion rate.
	In this work, we employ a divide-and-conquer approach to overcome these limitations by transforming the incident fields into uniform distributions that readily lend themselves for characterization by standard on-off detectors.
	Since the exact statistics of the light stream are obtained from the click statistics, our technique is freely scalable to accommodate -- in principle -- arbitrarily large photon fluxes.
	Our experiments pave the way towards genuine integrated photon-number-resolving detectors for advanced on-chip photonic quantum networks.
\end{abstract}

\date{\today}
\maketitle

	Quantum information science is at the cutting edge of modern physics and technology.
	In this context, perhaps the most ambitious goal is to realize scalable quantum information processing and computing based exclusively on linear optical configurations and photon-counting devices~\cite{1,2,3}.
	Notably, any such optical quantum-computing scheme hinges on the ability to detect and manipulate the states of light at the single-photon level:
	Quantum cryptography, entanglement swapping, and quantum teleportation, to name a few, would clearly be impossible without reliable single-photon-counting devices~\cite{4,5,6,7,8,9,10}.
	Moreover, exact photon counts provide access to genuine photon number statistics, and in turn are the principal means of reliably establishing the non-classicality of any type of light field~\cite{11,12,13,14,15,16,17,18}.
	Another potential application of photon-number-resolving detectors (PNRs) was recently highlighted in the context of coherent optical communications~\cite{19}, where they enable coherent optical communications with a performance superior to the standard quantum limit, even in the high mean photon number regime.

	Existing schemes for measurements at the single-photon level employ on-off detectors, e.g. avalanche photodiodes (APDs)~\cite{11}, and as such are inherently limited by the so-called dead time.
	When an APD is triggered, it typically remains ``blind'' for several nanoseconds thereafter, and as a result, succeeding photons impinging on the detector cannot be registered~\cite{16}.
	In addition to being detrimental to the overall detection efficiency, this effect may corrupt the very photon statistics one strives to characterize.
	Moreover, this saturation effect also introduces undesired correlations to the count sequences.
	To this date, the perhaps most noticeable scheme for PNRs is based on superconducting nanowires~\cite{20}.
	Yet, on many occasions, cryogenic measurements may be impractical, or the incident photon flux may exceed the capacity of the system.
	Evidently, a fundamentally different approach will be required to reconcile the demands for high speed, low noise, and maximized quantum efficiency with the ever increasing count rates required by modern technologies~\cite{21,22,23,24,25}.

	In contrast to PNRs, on-off detectors deliver well-defined ``clicks'' upon excitations with any non-zero number of photons~\cite{26}.
	Consequently, the by far most accessible quantum-optical measurements are click-counting statistics, instead of actual photon counts~\cite{27}.
	The question naturally arises as to whether it is possible to circumvent the limitations of on-off detectors, and to exploit these robust and widely available components to accurately measure multiple photons.

	In this work, we propose, implement and characterize a photon counting device based on a multiplexed array of on-off detectors.
	In our arrangement, the discrete evolution dynamics of the incident light field is manipulated so as to distribute the photons uniformly between the individual channels.
	Crucially, the click-counting statistics obtained from this type of multiplexed sensor converge to the actual photon-counting statistics as the number of independent detectors is increased~\cite{28}.
	Along these lines, it becomes possible to reliably probe the non-classicality of arbitrary optical input fields in a readily scalable and integrated fashion.

	\begin{figure}[ht]
		\centering
		\includegraphics*[width=7.5cm]{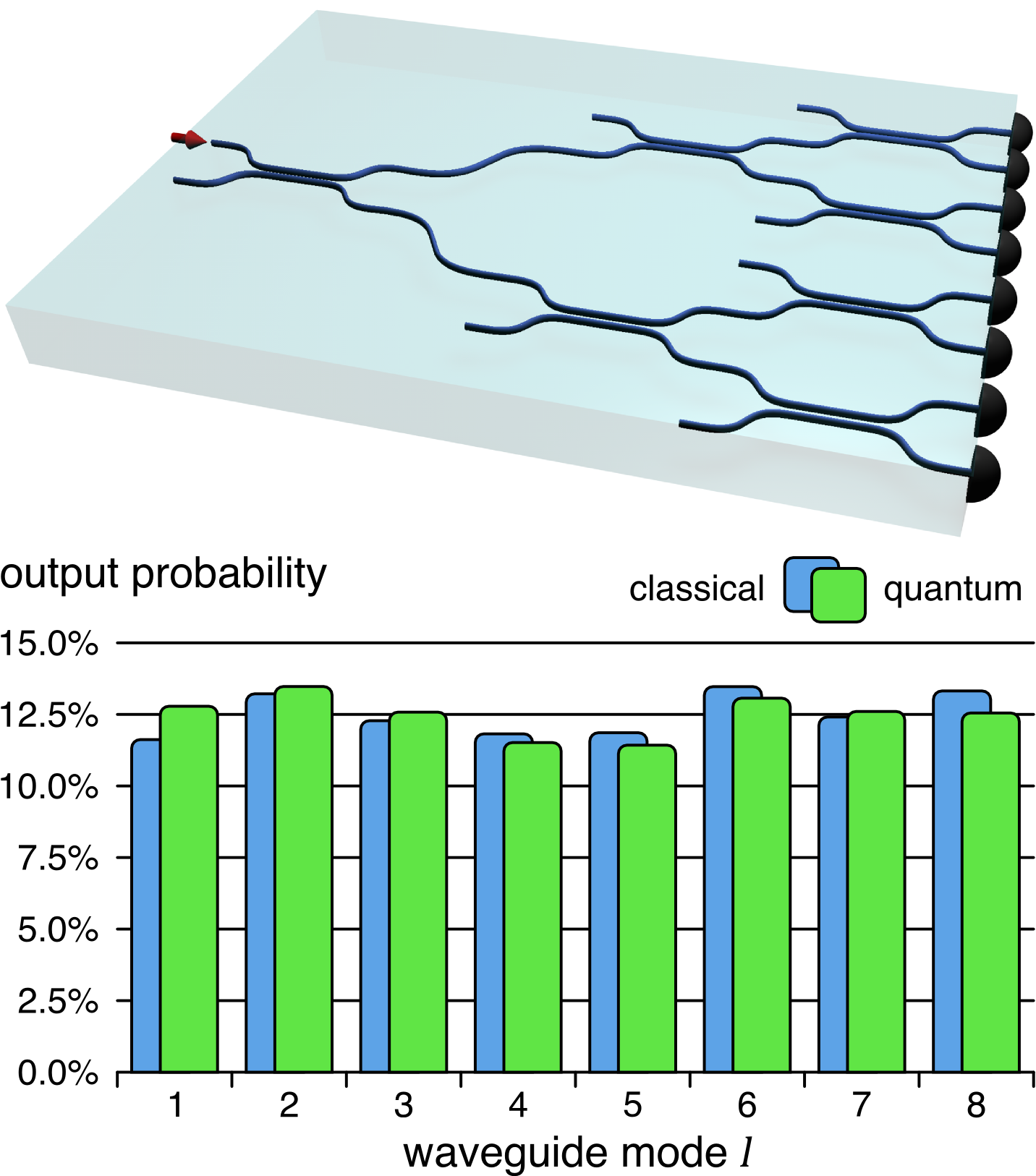}
		\caption{
			Top:
			Sketch of the 1-to-8 optical integrated multiplexer consisting of three beam splitting stages.
			The output fields are fed into APDs.
			Bottom:
			Output click statistic of the above multiplexer for classical attenuated laser light (blue), as well as single-photon Fock-states (green).
			In the experiment, both input states yield a flat uniform output statistic provided by the high quality of the optical-integrated device.
		}\label{fig:2}
	\end{figure}
	Let us first consider a stream of single-photon states being routed through a uniform 1-to-$N$ multiplexer and onto an array of APDs (see Fig.~\ref{fig:2}; top).
	In our method, this is achieved by cascading $m$ stages of 50/50 beam splitters, yielding $N=2^m$.
	Under these premises, every single-photon will have a probability of $1/N$ to be detected in one of the $N$ channels.
	Due to the spatially extended wave function, any two incoming photons are likely to be found in different outputs with a probability of $1-1/N$:
	When one photon is detected, the global probability of registering the next photon in any of the remaining APDs is $(N-1)$ times greater than in the same one.
	This remains true even if the two photons enter the system simultaneously. In this manner, the fidelity of the device improves with the number of output ports, and is even independent of the type of input state~\cite{28}.

	To experimentally demonstrate the functionality of our approach, we realized a discrete network of integrated 50/50 beam splitters cascaded in $m=3$ steps, yielding a total of $N=8$ output channels.
	These photonic structures were implemented in fused silica glass by means of the femtosecond laser writing technique~\cite{29,30}, see Appendix~\ref{app:Methods}.
	As input states we consider the two limiting cases:
	i) Low-intensity laser light, which can be idealized as a quantum coherent state~\cite{5}.
	As such, its statistics feature substantial yet temporally fluctuating bunching of photons, making it a perfect test case for the photon number resolving capabilities of our setup.
	ii) Heralded single photons from a spontaneous parametric down conversion source, which approximately represent the ideal scenario of single-photon Fock states~\cite{31}.

\begin{figure}[ht]
		\centering
		\includegraphics*[width=7.5cm]{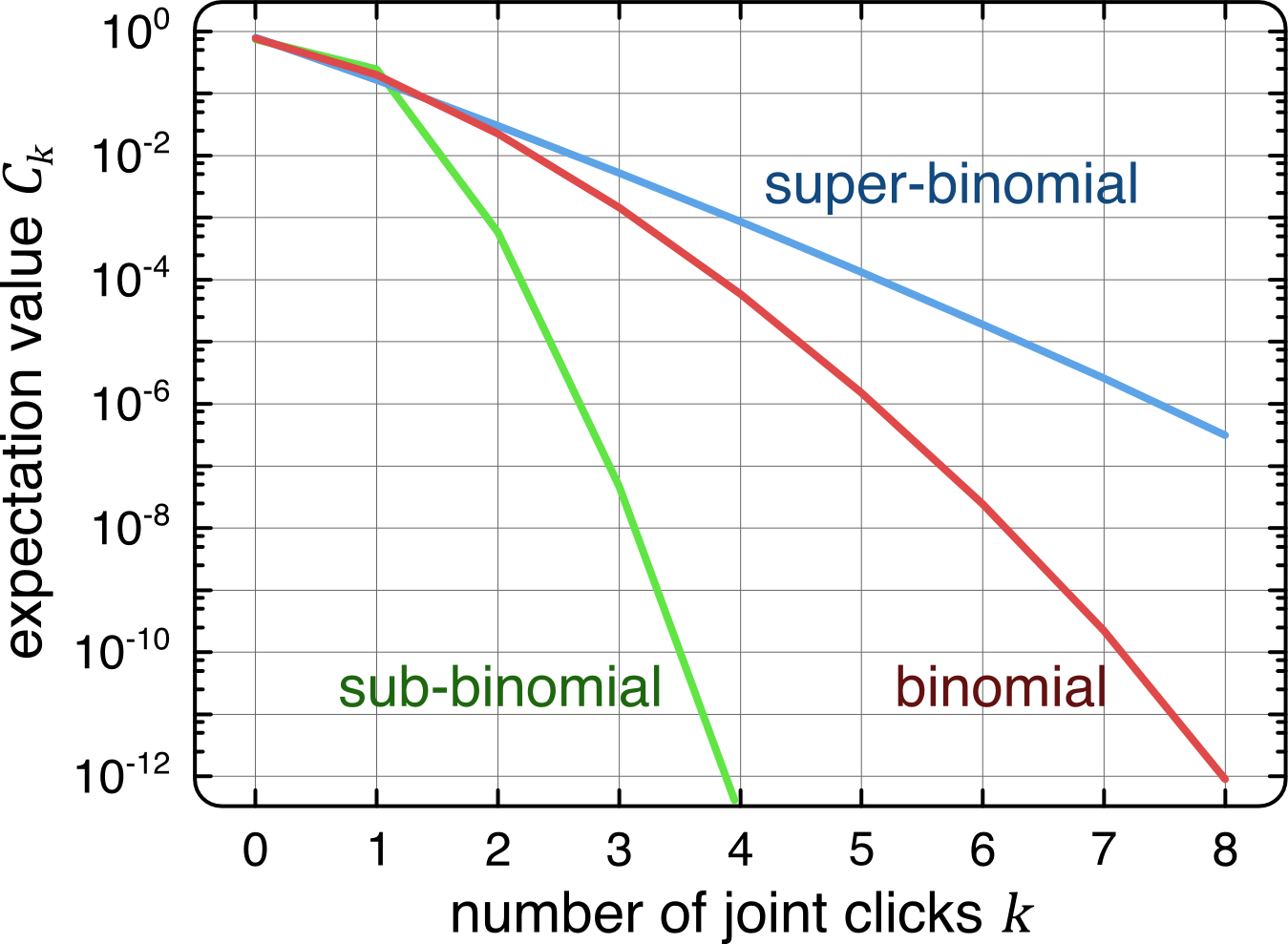}
		\caption{
			Theoretical click-counting statistics for three different photon number distributions at an average number of clicks of 0.25 per measurement.
			The resulting  values are $0$ for binomial, $-0.22$ for sub-binomial, and $+0.22$ for super-binomial click statistics.
		}\label{fig:1}
	\end{figure}

	In the multiplexer, classical coherent states $|\psi_{\rm in}\rangle=|\alpha\rangle$ are split into eight spatially separated coherent states of equal amplitude, i.e. $|\psi_{\rm out}\rangle=$
		$|{-}\alpha/\sqrt 8\rangle$
		$|{\rm i}\alpha/\sqrt 8\rangle$
		$|\alpha/\sqrt 8\rangle$
		$|{\rm i}\alpha/\sqrt 8\rangle$
		$|{-}\alpha/\sqrt 8\rangle$
		$|{\rm i}\alpha/\sqrt 8\rangle$
		$|{-}\alpha/\sqrt 8\rangle$
		$|{-}{\rm i}\alpha/\sqrt 8\rangle$.
	Consequently, a perfect photon counting characterization should yield a Poissonian photon number distribution.
	When one instead evaluates the counting click coincidences, it can be analytically shown that the resulting click-counting statistics have to follow a binomial distribution~\cite{28}.
	Note that in case of a sub- (or super-) Poissonian photon number distribution, it likewise follows that the click statistics are sub- (or super-) binomial, respectively (see for instance Fig.~\ref{fig:1}).

	In general, this behavior with respect to $N$ on-off detectors is mathematically described by the expectation value $C_k=\left\langle{:}\binom{N}{k}\hat\pi^k(\hat 1-\hat\pi)^{N-k}{:}\right\rangle$.
	Here, the normal ordering for the bosonic operators~\cite{32} is indicated by the symbol ${:}\ldots{:}$, the number of APDs that click in a certain time window is $0\leq k\leq N$, and the operator $\hat\pi={:}\hat 1-\exp[-(\eta\hat n+\nu)/N]{:}$ includes the photon number operator $\hat n$ and accounts for a realistic detector with quantum efficiency $\eta$ and dark counts $\nu$.
	In this expression, the exponential operator corresponds to the projector of the vacuum operator, and as expectation value yields the probability of zero clicks~\cite{33}.

	In order to quantify the binomial character of the click-counting statistics, we employ the parameter
	\begin{align}
		\label{eq:1}
		Q_B&=N\frac{\langle(\Delta k)^2\rangle}{\langle k\rangle(N-\langle k\rangle)}-1,\text{ with}
		\\\nonumber
		Q_B&\left\lbrace\begin{array}{ll}
			>0 &\text{ for super-binomial click statistics},\\
			=0 &\text{ for binomial click statistics},\\
			>0 &\text{ for sub-binomial click statistics},\\
		\end{array}\right.
	\end{align}
	where $\langle k\rangle=\sum_{k=0}^N k C_k$ and $\langle(\Delta k)^2\rangle=\langle k^2\rangle-\langle k\rangle^2$ represent the average number of clicks and the variance thereof, respectively~\cite{34}.
	For an ideal coherent state, the mean and the variance of the click-counting statistics are analytically calculated as $\langle k\rangle=Np$ and $\langle(\Delta k)^2\rangle=Np(1-p)$, respectively, with $p=1-\exp[-(\eta|\alpha|^2+\nu)/N]$.
	When substituting these two expressions into Eq.~\eqref{eq:1} we find that $Q_B=0$ regardless of the quantum efficiency and the dark count rate.
	Accordingly, it is a sufficient criterion to differentiate between classical and non-classical light:
	While super-binomial click statistics such as these of realistic laser light are characterized by $Q_B>0$, genuine quantum entities, such as Fock states, necessarily features $Q_B<0$.
	Note that in order for this parameter to be meaningful, the ensemble has to include $N\geq2$ on-off detectors, otherwise measurements of any input state would produce $Q_B=0$~\cite{34,35}.

	In a first set of experiments, we measured the absolute number of click coincidences $M_k$ and the relative frequencies $C^{\rm exp}_k=M_k/M$ for attenuated laser light, where $k$ again represents the number of clicks within a time window of $10\,{\rm ns}$ and $M=\sum_k M_k$ is the total number of time windows.
	Further details on our setup are given in the Appendix~\ref{app:Methods}.
	From these measurements, we extract a positive value $Q_B=(1.712\pm0.026)\cdot10^{-2}$, confirming that the click statistics is indeed super-binomial as expected for a classical coherent state.
	Beyond the $Q_B$ parameter, which relies solely on second-order correlations, one can also employ the higher-order correlations contained within the matrix of moments to identify non-classical behavior (see Appendix~\ref{app:SI}).

	In our second set of experiments, we used our device to characterize the fidelity of a heralded single-photon source based on spontaneous parametric down conversion.
	As in the previous case, the input state $|\psi_{\rm in}\rangle=\hat a^\dagger|0\rangle$ is spatially distributed and thus transformed according to the expression $|\psi_{\rm out}\rangle=(
		{-}\hat a_1^\dagger
		{+}{\rm i}\hat a_2^\dagger
		{+}\hat a_3^\dagger
		{+}{\rm i}\hat a_4^\dagger
		{-}\hat a_5^\dagger
		{+}{\rm i}\hat a_6^\dagger
		{-}\hat a_7^\dagger
		{-}{\rm i}\hat a_8^\dagger
		)|0\rangle/\sqrt{8}$,
	where $\hat a^\dagger_l$ represents the bosonic creation operators of the $l^{\rm th}$ waveguide mode.
	The high fidelity of our device is confirmed by the homogeneity of the single photon number output distribution with an average of $12.5\pm0.6$ (see Fig.~\ref{fig:2}; bottom).
	Measurement data clearly demonstrates the non-classicality of the input state, with $Q_B=-(2.002\pm0.049)\cdot10^{-2}<0$.

\begin{figure}[ht]
		\centering
		\includegraphics*[width=7.5cm]{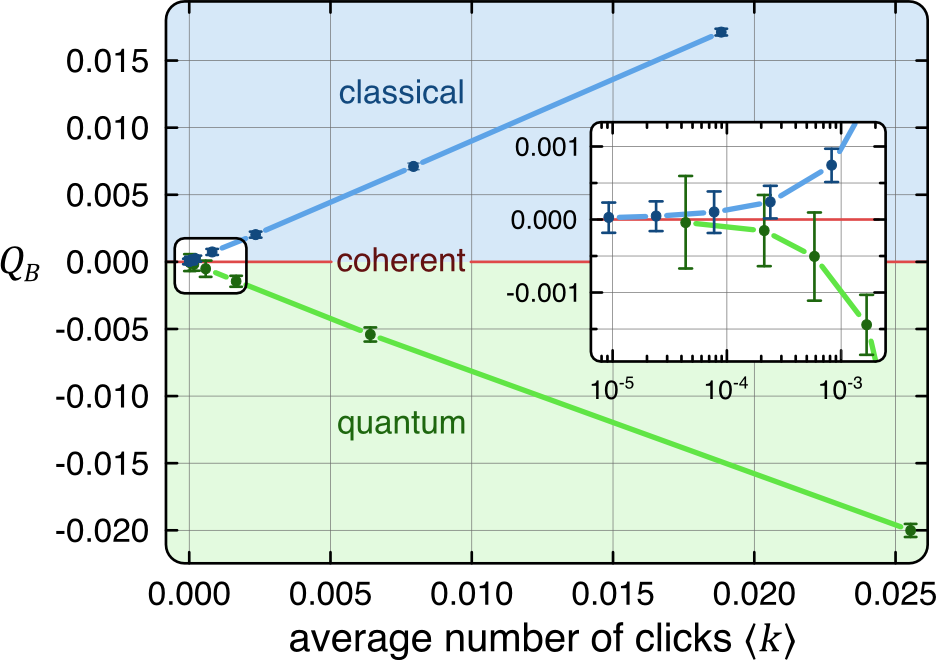}
		\caption{
			Experimentally obtained $Q_B$ for two different click statistics versus a variation of the average click number $\langle k\rangle$.
			The inset shows a semilogarithmic section for the smallest achieved $\langle k\rangle$s.
			An attenuated laser light (blue curve) is always accompanied by a positive $Q_B$ (super-binomial), whereas our single photon source (green curve) constantly shows a negative $Q_B$ (sub-binomial).
			In both cases, a coherent state (binomial photon number distribution) is covered within the error-bars at very low count rates.
		}\label{fig:3}
	\end{figure}

	Finally, we analyzed the dependence of the fidelity of our device for different source brightness levels.
	To this end, we determined the $Q_B$ parameter for classical laser light with different attenuation ratios, as well as for Fock states at different count rates.
	Figure~\ref{fig:3} illustrates that our scheme allows for a clear distinction between classical and non-classical behavior.
	Note that, as the incident intensity decreases, the chance of multiple photons entering the device in any given time slot gradually converges to zero, as does the corresponding $Q_B$ parameter.
	Whereas the sign of $Q_B$ remains well defined throughout this process, the uncertainty determined by the experimental conditions [see Appendix~\ref{app:Methods}] eventually exceeds the absolute value.
	Nevertheless, standard quantum sources routinely feature count rates well above this limit, placing them firmly within the window of confidence of our characterization scheme.

	In conclusion, we have introduced a new paradigm for integrated photon-number resolved measurements based on distributed sensing with multiplexed arrays of conventional on-off detectors.
	In this divide-and-conquer approach, the limiting factor of detector dead time is overcome by transforming the incident fields into extended uniform distributions.
	As such, coincidences in the same measurement channel are reliably suppressed even for highly multi-photon input states.
	Measuring the click statistics of the detector ensemble therefore provides the means to determine the actual photon-counting statistics.
	Consequently, our technique can in principle be scaled to allow for arbitrarily high numbers of incident photons, irrespective of the dead time of the individual detectors used.
	On a higher level, similar multiplexing schemes could even be employed to boost the performance of PNR detectors relying on other mechanisms of registration.
	Our experiments pave the way towards genuine integrated photon-number-resolving detectors for advanced on-chip photonic quantum networks.

\appendix

\section{Methods}\label{app:Methods}
	\subsection{Fabrication \& characterization}
	The multiplexing waveguide network was fabricated by means of the direct femtosecond laser inscription in fused silica glass~\cite{29,30}.
	It is designed to match the technical standards of the attached V-groove fiber array with $127\,{\rm \mu m}$ pitch which collects the photon outcome and feeds it into single-photon click-detectors.
	In order to inject weak laser light, different neutral density (ND) filters were placed into the beam of a laser diode emitting at $808\,{\rm nm}$.
	The attenuated light was coupled into a single mode (SM) polarization maintaining (PM) fiber attached to the input wave guide of the multiplexer.
	For the other set of experiments, single photons of $815\,{\rm nm}$ were generated by spontaneous parametric down conversion (SPDC) in a BiB${}_3$O${}_6$ crystal and coupled into two SM and PM fibers.
	One of the fibers was directly connected to a single-photon detector to provide a herald for the other twin photon, which was delivered directly to the device's injection site.
	By using the second photon as trigger, we can suppress any noticeable influence of dark counts.
	For reasons of experimental convenience, the overall photon flux was again controlled by an appropriate choice of ND filters.

	\subsection{Measurement methods \& errors}
	The photon clicks by our APDs were collected by a time tagging card capable to handle up to 16 detectors at the same time.
	The coincidence time window, wherein two or more clicks are interpreted as a joint event, was set to $\Delta\tau=10\,{\rm ns}$ for all measurements.
	In turn, the number of measurements for attenuated laser light was determined by $T/\Delta\tau$ where $T$ is the overall recording time.
	In contrast, the number of quantum measurements was determined by the photon flux of the trigger photons, and therefore remains independent of the coincidence time window.
	In both cases, the overall measurement time was chosen in a way to collect at least 100 million non-zero click events, ranging from 44 seconds up to more than 60 hours.

	The accuracy of the calculated $Q_B$ parameter is chiefly dependent on the number of measurements.
	In an ideal set of data, the frequencies $C_k$ would be distributed according to the graphs shown in Fig.~\ref{fig:1}.
	If a measurement is terminated before the highest number of joint clicks yields a non-zero quantity, it cannot perfectly match the (sub- or super-) binomial distribution.
	Consequently, the extracted $Q_B$ parameter is subject to a systematic uncertainty.
	A second source of errors results from imperfections in the splitting ratio of the multiplexing device, corresponding to small deviations from the ideal case of homogeneously distributed outputs (see Fig.~\ref{fig:2}).
	This also influences the coincidence click statistics and, hence, the error for the $Q_B$ parameter.
	In turn, the resulting overall range of error defines a certain minimum count rate.

	We would like to emphasize that our method is capable of reliably identifying the signature of the input state, whether classical or quantum (see Fig.~\ref{fig:3}), without the need for any corrections or post processing of raw data.

\section{High-order correlations}\label{app:SI}
	In order to analyze high order correlations of the light fields under study, we use the approach recently introduced in~\cite{33}.
	Along these lines, we compute the matrix of moments $MoM$ directly from the experimental click-counting statistics $C_k^{\rm exp}$.
	This is done using the expression $MoM=(\langle{:}\hat\pi^{m+n}{:}\rangle)_{m,n}$.
	Here,
	\begin{align}
		\label{eq:S1}
		&\langle{:}\hat\pi^{m+n}{:}\rangle
		\\\nonumber
		=&\frac{(N{-}[m{+}n])!}{N!}\sum_{k=0}^N k(k{-}1)\ldots(k{-}[m{+}n]{+}1)C_k^{\rm exp},
	\end{align}
	with the indices $m,n$ ranging from $0$ to $\lfloor N/2\rfloor$, where $\lfloor\ldots\rfloor$ represents the floor function.
	It can be shown that the second principal minors of the $MoM$ can be expressed in terms of the $Q_B$ parameter, and as a result for classical states the associated $MoM$ is expected to be non-negative.
	The positive semi-definiteness of the $MoM$ can be established through the analysis of corresponding eigenvectors and eigenvalues.
	In other words, the $MoM$ will be positive definite if all its eigenvalues are positive.
	Therefore, if we find at least one negative eigenvalue, we can unequivocally conclude that the light field being characterized is non-classical.
	This condition is mathematically described by
	\begin{align}
		\label{eq:S2}
		0> \vec f^\dagger{MoM}\vec f=\langle {:}\hat f^\dagger\hat f{:}\rangle=\sum_{m,m'=0}^{\lfloor N/2\rfloor} f^\ast_mf_{m'} \langle {:}\hat\pi^{m+m'}{:}\rangle,
	\end{align}
	where we have defined the operators $\hat f=f_0+f_1\hat\pi+\ldots+f_{\lfloor N/2\rfloor}\hat\pi^{\lfloor N/2\rfloor}$ and $\vec f=(f_0,\ldots,f_{\lfloor N/2\rfloor})^T$ represents the eigenvectors of $MoM$.
	For a classical coherent source, the moments can be calculated analytically using the expression $\langle{:} \hat\pi^m{:}\rangle=p^m=(1-\exp[-(\eta|\alpha|^2+\nu)/N])^m$.
	This indicates that the  of an ideal coherent state exhibits one positive eigenvalue given by $(1-p^{2\lfloor N/2\rfloor})/(1-p^2)$, and a total of $\lfloor N/2\rfloor$ zero eigenvalues for $N\geq2$.

	Here, we consider a 1-to-8 multiplexer device, which corresponds to a $5\times5$ $MoM$:
	\begin{align}
		\label{eq:S3}
		MoM=\begin{pmatrix}
			\langle {:}\hat\pi^{0}{:}\rangle & \cdots & \langle {:}\hat\pi^{4}{:}\rangle \\
			\vdots & \ddots & \vdots \\
			\langle {:}\hat\pi^{4}{:}\rangle & \cdots & \langle {:}\hat\pi^{8}{:}\rangle
		\end{pmatrix}.
	\end{align}
	Using the measured $C^{\rm exp}_k$, we numerically obtain the eigenvalues and eigenvectors.
	The results of this analysis for experimental data obtained over $83.2\,{\rm s}$ (corresponding to  windows) are shown in Tables~\ref{tab:SI} and~\ref{tab:SII} for the classical and quantum states, respectively, whereas Figure~\ref{fig:3} shows a graphical representation of these quantities.

	\begin{table*}[ht]
		\begin{tabular}{|c|c|c|c|}
			\hline
			\hspace*{0.25cm} Eigenvalues\hspace*{0.25cm} & $\langle {:}\hat f^\dagger\hat f{:}\rangle$ & $\sigma(\langle {:}\hat f^\dagger\hat f{:}\rangle)$ & $\Sigma$ \\
			\hline
			1 & $1\cdot{10}^{00} \pm 3\cdot {10}^{-10}$ & $3.2936\cdot{10}^{-10}$ & $3.03619\cdot{10}^{9}$\\
			\hline
			2 & $1.40231\cdot{10}^{-06} \pm 3\cdot{10}^{-09}$ & $3.88066\cdot{10}^{-9}$ & 361.359\\
			\hline
			3 & $4.3309\cdot{10}^{-13} \pm 3\cdot{10}^{-12}$ & $4.43711\cdot{10}^{-12}$ & 0.0976064\\
			\hline
			4 & $-8.59786\cdot{10}^{-16} \pm 3\cdot{10}^{-15}$ & $8.87048\cdot{10}^{-15}$ & 0.0969266\\
			\hline
			5 & $-2.47982\cdot{10}^{-22} \pm 3\cdot{10}^{-22}$ & $6.09801\cdot{10}^{-22}$ & 0.406661\\
			\hline
		\end{tabular}
		\caption{
			Correlation measurements for classical coherent light.
			As expected from a classical source no one of the eigenvalues is significantly negative.
		}\label{tab:SI}
	\end{table*}
	\begin{table*}[ht]
		\begin{tabular}{|c|c|c|c|}
			\hline
			\hspace*{0.25cm} Eigenvalues\hspace*{0.25cm} & $\langle {:}\hat f^\dagger\hat f{:}\rangle$ & $\sigma(\langle {:}\hat f^\dagger\hat f{:}\rangle)$ & $\Sigma$ \\
			\hline
			1 & $1.00001\cdot{10}^{00} \pm 3\cdot{10}^{-10}$ & $7.77014\cdot{10}^{-9}$ & $1.287\cdot{10}^{8}$\\
			\hline
			2 & $-1.38281\cdot{10}^{-05} \pm 3\cdot{10}^{-09}$ & $1.6425\cdot{10}^{-8}$ & 841.894\\
			\hline
			3 & $-3.78528\cdot{10}^{-14} \pm 3\cdot{10}^{-12}$ & $4.47567\cdot{10}^{-14}$ & 0.845747\\
			\hline
			4 & $2.78037\cdot{10}^{-21} \pm 3\cdot{10}^{-15}$ & $3.29078\cdot{10}^{-21}$ & 0.844898\\
			\hline
			5 & $0 \pm 0$ & $0$ & {\sl Indetermined}\\
			\hline
		\end{tabular}
		\caption{
			Correlation measurements for the single-photon Fock states.
		}\label{tab:SII}
	\end{table*}

	The significance for these experimentally obtained eigenvalues can be quantified using the expression $\Sigma=\frac{|\langle {:}\hat f^\dagger\hat f{:}\rangle|}{\sigma(\langle {:}\hat f^\dagger\hat f{:}\rangle)}$, where $\sigma(\langle {:}\hat f^\dagger\hat f{:}\rangle)$ is the standard error of the mean $\langle {:}\hat f^\dagger\hat f{:}\rangle$.
	Note that the two negative eigenvalues, which unexpectedly occur in the classical case, can be considered zero since their absolute values fall within the error margin.
	The values of $\Sigma$ for all measurements are shown in the fourth column of Table~\ref{tab:SI}, illustrating that all eigenvalues are non-negative within the range of their significance.
	A similar validity check was also applied for our measurements of single photon Fock-states (Table~\ref{tab:SII}).
	As in the previous case, we calculate the significance of the deviation from the classical bound, i.e. zero. This is done using the second eigenvector $\vec f=($
		$0.00384503$,
		${-}0.999993$,
		${-}0.000251841$,
		$-5.430982775\cdot10^{-8}$,
		$0)^T$
	which determines the operator test.
	The obtained value of $\Sigma=841.894$ certifies that the state under consideration is indeed non-classical.

\section*{Acknowledgments}
	Financial support by the German Ministry of Education and Research (Center for Innovation Competence programme, grant no. 03Z1HN31), the Thuringian Ministry for Education, Science and Culture (Research group Spacetime, grant no. 11027-514) and Deutsche Forschungsgemeinschaft (grant no. NO462/6-1 and SFB 652) is gratefully acknowledged.
	M. Heinrich was supported by the German National Academy of Sciences Leopoldina (grants no. LPDS 2012-01 and LPDR 2014-03).


\begin{thebibliography}{100}
	\bibitem{1} Kok, P., Munro, W. J., Nemoto, K., Ralph, T. C., Dowling, J. P. and Milburn, G. J., ``Linear optical quantum computing with photonic qubits,'' Rev. Mod. Phys. {\bf 79}, 135 (2007).
 	\bibitem{2} Knill, E., Laflamme, R. and Milburn, G. J. ``A scheme for efficient quantum computation with linear optics,''  Nature {\bf 409}, 46-52 (2000).
	\bibitem{3} Humphreys, P. C., Metcalf, B. J., Spring, J. B., Moore, M., Jin, X-M., Barbieri, M., Kolthammer, W. S. and Walmsley, I. A. ``Linear Optical Quantum Computing in a Single Spatial Mode,'' Phys. Rev. Lett. {\bf 111}, 150501 (2013).
	\bibitem{4} Gr\"afe, M., Heilmann, R., Perez-Leija, A., Keil, R., Dreisow, F., Heinrich, M., Moya-Cessa, H., Nolte, S., Christodoulides, D. N. and Szameit, A. ``On-chip generation of high-order single-photon W-states,'' Nature Photonics {\bf 8}, 791-795 (2014).
	\bibitem{5} Afek, I., Ambar, O. and Silberberg, Y. ``High-NOON states by mixing quantum and classical light,'' Science {\bf 328}, 879-881 (2010).
	\bibitem{6} Morin, O., Bancal, J-D., Ho, M., Sekatski, P., D'Auria, V., Gisin, N., Laurat, J. and Sangouard, N. ``Witnessing trustworthy single-photon entanglement with local homodyne measurements,'' Phys. Rev. Lett. {\bf 110}, 130401 (2013).
	\bibitem{7} Papp, S. B., Choi, K.S., Deng, H., Lougovski, P., Van Enk, S. J. and Kimble, H. J., ``Characterization of multipartite entanglement for one photon shared among four optical modes,'' Science {\bf 324}, 764-768 (2009).
	\bibitem{8} Kimble, H. J. ``The quantum internet,'' Nature {\bf 453}, 1023-1030 (2008).
	\bibitem{9} Silberhorn, C. ``Detecting quantum light,'' Contemp. Phys. {\bf 48}, 143 (2007).
	\bibitem{10} Lita, A. E., Miller, A. J. and  Nam, S. W. ``Counting near-infrared single-photons with 95\% efficiency,'' Opt. Express {\bf 16}, 3032 (2008).
	\bibitem{11} Chrapkiewicz, R. ``Photon counts statistics of squeezed and multimode thermal states of light on multiplexed on-off detectors,'' JOSA B {\bf 31}, B8-B13 (2014).
	\bibitem{12} Hadfield, R. H. ``Single-photon detectors for optical quantum information applications,'' Nature Photonics {\bf 3}, 696-705 (2009).
	\bibitem{13} Allevi, A. and Bondani, M. ``Statistics of twin-beam states by photon-number resolving detectors up to pump depletion,'' JOSA B {\bf 31}, B14-B19 (2014).
	\bibitem{14} Achilles, D. , Silberhorn, C., Sliwa, C., Banaszek, K. and Walmsley, I. A. ``Fiber assisted detection with photon-number resolution,'' Opt. Lett. {\bf 28}, 2387-2389 (2003).
	\bibitem{15} Levine, Z. H., Glebov, B. L., Migdall, A. L., Gerrits, T., Calkins, B., Lita, A. E. and Nam, S. W. ``Photon-number uncertainty in a superconducting transition edge sensor beyond resolved-photon-number determination,'' JOSA B {\bf 31}, B20-B24 (2014).
	\bibitem{16} Kalashnikov, D. and Krivitsky, L. ``Measurement of photon correlations with multipixel photon counters,'' JOSA B {\bf 31}, B25-B33 (2014).
	\bibitem{17} O'Brien, J. L. ``Optical quantum computing,'' Science {\bf 318}, 1567-1570 (2007).
	\bibitem{18} Sridhar, N., Shahrokhshahi, R., Miller, A. J., Calkins, B., Gerrits, T., Lita, A., Nam, S. W. and Pfister, O. ``Direct measurement of the Wigner function by photon-number-resolving detection,'' JOSA B {\bf 31}, B34-B40 (2014).
	\bibitem{19} Becerra, F. E., Fan, J. and Migdall, A. ``Photon number resolution enables quantum receiver for realistic coherent optical communications,'' Nature Photonics {\bf 9}, 48-53 (2015).
	\bibitem{20} Divochiy, A., Marsili, F., Bitauld, D., Gaggero, A., Leoni, R., Mattioli, F., Korneev, A., Seleznev, V., Kaurova, N., Minaeva, O., Gol'tsman, G., Lagoudakis, K. G., Benkhaoul, M., L\'evy, F. and Fiore, A. ``Superconducting nanowire photon-number-resolving detector at telecommunication wavelengths,'' Nature Photonics {\bf 2}, 302-306 (2008).
	\bibitem{21} Fujiwara, M. and Sasaki, M. ``Direct measurement of photon number statistics at telecom wavelengths using a charge integration photon detector,'' Appl. Opt. {\bf 46}, 3069-3074 (2007).
	\bibitem{22} Gol'tsman, G. N., Okunev, O., Chulkova, G., Lipatov, A., Semenov, A., Smirnov, K., Voronov, B., Dzardanov, A., Williams, C. and Sobolewski, R. ``Picosecond superconducting single-photon optical detector,'' Appl. Phys. Lett. {\bf 79}, 705 (2001).
	\bibitem{23} Dauler, E. A., Robinson, B. S., Kerman, A. J., Yang, J. K. W., Rosfjord, K. M., Anant, V., Voronov, B., Gol'tsman, G. and Berggren, K. K. ``Multi-Element Superconducting Nanowire Single-Photon Detector,'' IEEE Trans. Appl. Superconductivity {\bf 17}, 279-284 (2007).
	\bibitem{24} Korneev, A., Kouminov, P., Matvienko, V., Chulkova, G., Smirnov, K., Voronov, B., Gol'tsman, G. N., Currie, M., Lo, W., Wilsher, K., Zhang, J., S\l{}ysz, W., Pearlman, A., Verevkin, A. and Sobolewski, R. ``Sensitivity and gigahertz counting performance of NbN superconducting single-photon detectors,'' Appl. Phys. Lett. {\bf 84}, 5338-5340 (2004).
	\bibitem{25} Robinson, B. S., Kerman, A. J., Dauler, E. A., Barron, R. J., Caplan, D. O., Stevens, M. L., Carney, J. J., Hamilton, S. A., Yang, J. K. W. and Berggren, K. K. "781-Mbit/s photon-counting optical communications using a superconducting nanowire detector",  Opt. Lett. {\bf 31}, 444-446 (2006).
	\bibitem{26} Zambra, G., Andreoni, A., Bondani, M., Gramegna, M., Genovese, M., Brida, G., Rossi, A. and Paris, M. G. A. ``Experimental Reconstruction of Photon Statistics without Photon Counting,'' Phys. Rev. Lett. {\bf 95}, 063602 (2005).
	\bibitem{27} Luis, A., Sperling, J. and Vogel, W. ``Nonclassicality phase-space functions: more insight with less detectors,'' arXiv: 1412.3826. 
	\bibitem{28} Sperling, J., Vogel, W. and Agarwal, G. S. ``True photocounting statistics of multiple on-off detectors,''  Phys. Rev. A {\bf 85}, 023820 (2012).
	\bibitem{29} Davis, K. M., Miura, K., Sugimoto, N. and Hirao, K. ``Writing waveguides in glass with a femtosecond laser,'' Opt. Lett. {\bf 21}, 1729-1731 (1996).
	\bibitem{30} Marshall, G. D, Politi, A., Matthews, J. C. F., Dekker, P., Ams, M., Withford, M. J. and O'Brien, J. L. ``Laser written waveguide photonic quantum circuits,'' Opt. Express {\bf 17}, 12546-12554 (2009).
	\bibitem{31} Lvovsky, A. I., Hansen, H., Aichele, T., Benson, O., Mlynek, J. and Schiller, S. ``Quantum State Reconstruction of the Single-Photon Fock State,'' Phys. Rev. Lett. {\bf 87}, 050402 (2001).
	\bibitem{32} Vogel, W. and Welsch, D.-G. {\it Quantum Optics} (Wiley-VCH, Weinheim, 2006), 3rd ed.  
	\bibitem{33} Sperling, J., Vogel, W. and Agarwal, G. S. ``Correlation measurements with on-off detectors,'' Phys. Rev. A {\bf 88}, 043821 (2013).
	\bibitem{34} Sperling, J., Vogel, W. and Agarwal, G. S. ``Sub-Binomial Light,'' Phys. Rev. Lett. {\bf 109}, 093601 (2012).
	\bibitem{35} Bartley, T. J., Donati, G., Jin, X.-M., Datta, A., Barbieri, M. and Walmsley, I. A. ``Direct Observation of Sub-Binomial Light,'' Phys. Rev. Lett. {\bf 110}, 173602 (2013).
\end{thebibliography}
\end{document}